\documentclass[letterpaper]{aa}
\usepackage{graphicx}
\usepackage{natbib}
\usepackage{txfonts}

\bibpunct{(}{)}{;}{a}{}{,}

\begin{document}

   \title{Far-Infrared detection of methylene
          \thanks{Based on observations with ISO, an ESA project with instruments funded by ESA Member States (especially  the PI countries: France, Germany, the Netherlands and the United Kingdom) with the participation of ISAS and NASA.}
   }

   \author{ E. T. Polehampton
           \inst{1}
	   \and
	   K. M. Menten
           \inst{1}
           \and
           S. Br\"{u}nken
	   \inst{2}
           \and
	   G. Winnewisser
	   \inst{2}
	   \and
           J.-P. Baluteau
	   \inst{3}
            }

   \offprints{E. Polehampton:  \email{epoleham@mpifr-bonn.mpg.de}}

   \institute{
   Max-Planck-Institut f\"{u}r Radioastronomie, Auf dem H\"{u}gel 69, D-53121 Bonn, Germany 
   \and 
   Physikalisches Institut, Universit\"{a}t zu K\"{o}ln, D-50937 Cologne, Germany 
   \and
   Laboratoire d'Astrophysique de Marseille, CNRS \& Universit\'e de Provence, BP 8, F-13376 Marseille Cedex 12, France }

   \date{Received  / accepted }

 \abstract{We present a clear detection of CH$_{2}$ in absorption towards the molecular cloud complexes \object{Sagittarius B2} and \object{W49~N} using the ISO Long Wavelength Spectrometer. These observations represent the first detection of its low excitation rotational lines in the interstellar medium. Towards Sagittarius B2, we detect both ortho and para transitions allowing a determination of the total CH$_{2}$ column density of $N(\rm{CH_{2}})=(7.5\pm1.1)\times10^{14}$~cm$^{-2}$. We compare this with the related molecule, CH, to determine [CH/CH$_{2}]=2.7\pm0.5$. Comparison with chemical models shows that the CH abundance along the line of sight is consistent with diffuse cloud conditions and that the high [CH/CH$_{2}$] ratio can be explained by including the effect of grain-surface reactions.

\keywords{Infrared: ISM -- ISM: molecules -- Molecular data -- ISM: individual objects: Sagittarius~B2 -- ISM: individual objects: W49} }

\titlerunning{FIR detection of methylene}
\maketitle
%
%

\section{Introduction}

Methylene (CH$_{2}$) is thought to be a relatively abundant molecule in diffuse as well as in dense interstellar clouds, with similar abundances to CH \citep[e.g.][]{vandishoeck86,lee}. However, it has proved very difficult to detect observationally, mainly due to the inaccessibility of its rotational lines from the ground.

Methylene was initially proposed to explain an unidentified ultraviolet (UV) band observed in comets \citep{1942RvMP...14..195H,1942ApJ....96..314H} but later this band was shown to be associated with C$_{3}$ \citep{douglas}. CH$_{2}$ remains undetected in comets, but recently, several of its electronic bands have been observed in the interstellar medium (ISM) by \citet{lyu} using the Hubble Space Telescope. They tentatively detected CH$_{2}$ absorption bands in the UV spectrum towards two stars, HD154368 and $\zeta$~Oph.

Interstellar searches for rotational emission/absorption are hampered by the peculiar spectrum of CH$_{2}$, caused by its lightness and $b$-type selection rules \citep{michael}. This results in lines at widely varying wavelengths, all of which are either completely unobservable from the ground or are in difficult spectral regions that are at the edges of atmospheric windows or for which there are few telescopes with suitable instrumentation to observe them.

\citet{hollis} have, so far, made the only unambiguous identification of CH$_2$ in the ISM, confirming their earlier detection \citep{1989ApJ...346..794H}. They clearly identified the $4_{04}$--3$_{13}$ rotational transition with simultaneous measurements of multiple fine-structure features between 68 and 71~GHz. These were detected in emission towards \object{Orion~KL} and \object{W51~M}, both dense ``hot core'' sources, which provide excitation to the $4_{04}$ level which is $\approx 215$ K above the ground state. The lines occur in a rarely observed spectral region, close to the telluric 55--65~GHz O$_2$ bands. Accurate frequencies for this transition were measured by \citet{lovas}.

Low-excitation rotational transitions occur in the far-infrared (FIR) region, accessible by the Infrared Space Observatory (ISO) satellite (and in the future by SOFIA and Herschel), and around 320~$\mu$m (940 GHz), a spectral region just being opened from the ground. Although limited laboratory data on the rotational spectrum of CH$_{2}$ have existed for many years \citep{lovas,sears}, much more accurate frequencies are currently being measured at the Cologne Laboratory for Molecular Spectroscopy \citep{michael,bruenken} 

We have used these new data to search for low-lying rotational transitions of CH$_{2}$ observed by the ISO Long Wavelength Spectrometer \citep[LWS; ][]{clegg}. We clearly detect the strongest fine structure components of transitions from the lowest energy level of both ortho and para CH$_{2}$ towards \object{Sagittarius~B2} (\object{Sgr~B2}) and of ortho CH$_{2}$ towards \object{W49~N}. Both \object{Sgr~B2} and \object{W49} are giant molecular cloud complexes emitting strong FIR continuum spectra. This makes them particularly good targets for detecting absorption lines from intervening interstellar matter.

\object{Sgr~B2} was observed as part of a wide spectral survey using the LWS Fabry-P\'{e}rot (FP) mode, allowing us to search for transitions from all the other low-lying energy levels, as well as to compare the data with absorption from the chemically related CH molecule. We use the strong CH lines as a template to fit and calculate column densities for CH$_{2}$. In Sects. 4, 5 and 6 we present the results towards \object{Sgr~B2} and in Sect.~7 detail the results for \object{W49~N} and several other sources where CH$_{2}$ was not detected. We then compare the results with chemical models in Sect.~8.

\section{CH$_{2}$ rotational spectrum}

\begin{figure}
\resizebox{\hsize}{!}{\includegraphics{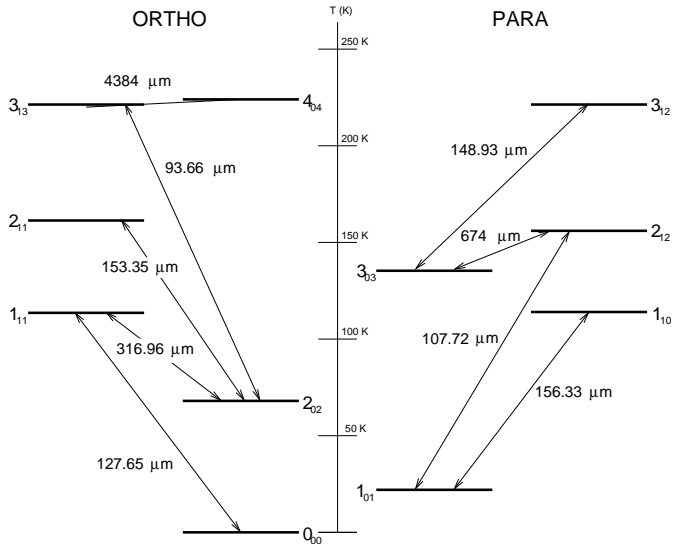}}
\caption{\label{levels}Low-lying rotational levels of the CH$_{2}$ molecule showing the wavelength of the strongest fine structure component for each transition.}
\end{figure}

Figure~\ref{levels} shows the low-lying rotational states of CH$_{2}$. The energy levels are denoted by $N_{KaKc}$, where $N$ is the rigid-body rotational quantum number. Each of these levels is split by electron spin-spin and spin-rotation interactions into three fine-structure levels, with the quantum number for total angular momentum excluding nuclear spin, $J$, equal to $N+1$, $N$, $N-1$ for $N>0$ ($J$=1 at $N$=0). Due to the presence of two protons with opposite nuclear spin, CH$_{2}$ has both ortho (nuclear spin quantum number, $I$=1) and para ($I$=0) forms. In the ortho levels, the non-zero nuclear spin interacts with the electrons to cause a further hyperfine splitting with total angular momentum quantum number, $F$=$J+1$, $J$, $J-1$. Selection rules for rotational transitions are $\Delta{J}$=0, $\pm$1 and $\Delta{F}$=0, $\pm$1.

\begin{table}[!t]
\caption{\label{freqs}Calculated wavelengths, Einstein coefficients ($A_{ij}$) and line strengths ($S_{ij}$) for the low-lying transitions of CH$_{2}$ (ignoring hyperfine-structure in the ortho transitions). The estimated error corresponding to the last significant figure of the calculated wavelengths is given in parenthesis. The Einstein coefficients and line strengths calculated here are half those given by \citet{sears}. The detected lines towards \object{Sgr B2} are shown in bold.}
\leavevmode \footnotesize
\begin{center}
\begin{tabular}[!t]{cccc}
\hline
\hline
Transition                   & Wavelength  & A$_{ij}$   & S$_{ij}$    \\ 
                             &  ($\mu$m)   & (s$^{-1}$) &  \\  
\hline 
1$_{11}$--0$_{00}$  $J$=0--1 & 127.31450 (6) &   0.0163 &     0.989\\
{\bf 1$_{{\bf 11}}$--0$_{{\bf 00}}$  ${ J}$=1--1} & {\bf 127.85823 (4)} &   {\bf 0.0163} &     {\bf 3.02}\\
{\bf 1$_{{\bf 11}}$--0$_{{\bf 00}}$  ${ J}$=2--1} & {\bf 127.64614 (5)} &   {\bf 0.0163} &     {\bf 4.99}\\
\hline
2$_{11}$--2$_{02}$  $J$=1--1 & 153.102312 (6) &   0.0105 &     3.31\\
2$_{11}$--2$_{02}$  $J$=1--2 & 154.302721 (7) &   0.00351 &     1.14\\
2$_{11}$--2$_{02}$  $J$=2--1 & 152.621311 (6) &   0.00217 &     1.14\\
2$_{11}$--2$_{02}$  $J$=2--2 & 153.814187 (6) &   0.00962 &     5.15\\
2$_{11}$--2$_{02}$  $J$=2--3 & 152.992296 (6) &   0.00216 &     1.14\\
2$_{11}$--2$_{02}$  $J$=3--2 & 154.178941 (6) &   0.00151 &     1.14\\
2$_{11}$--2$_{02}$  $J$=3--3 & 153.352914 (7) &   0.0125 &     9.25\\
\hline
3$_{13}$--2$_{02}$  $J$=2--1 &  93.5838 (1) &   0.0298 &     3.60\\
3$_{13}$--2$_{02}$  $J$=2--2 &  94.0309 (1) &   0.00559 &     0.685\\
3$_{13}$--2$_{02}$  $J$=2--3 &  93.7231 (2) &   0.000170 &     0.0205\\
3$_{13}$--2$_{02}$  $J$=3--2 &  93.7025 (2) &   0.0317 &     5.37\\
3$_{13}$--2$_{02}$  $J$=3--3 &  93.3967 (2) &   0.00392 &     0.657\\
3$_{13}$--2$_{02}$  $J$=4--3 &  93.6621 (2) &   0.0356 &     7.75\\
\hline
1$_{10}$--1$_{01}$  $J$=0--1 & 156.43947 (1) &   0.0133 &     0.500\\
1$_{10}$--1$_{01}$  $J$=1--0 & 155.66632 (1) &   0.00462 &     0.513\\
1$_{10}$--1$_{01}$  $J$=1--1 & 157.56539 (1) &   0.00318 &     0.366\\
1$_{10}$--1$_{01}$  $J$=1--2 & 156.76825 (1) &   0.00548 &     0.622\\
1$_{10}$--1$_{01}$  $J$=2--1 & 157.12503 (2) &   0.00323 &     0.615\\
1$_{10}$--1$_{01}$  $J$=2--2 & 156.33233 (1) &   0.0101 &     1.88\\
\hline
2$_{12}$--1$_{01}$  $J$=1--0 & 107.2947 (2) &   0.0134 &     0.487\\
2$_{12}$--1$_{01}$  $J$=1--1 & 108.1935 (2) &   0.0103 &     0.385\\
2$_{12}$--1$_{01}$  $J$=1--2 & 107.8170 (2) &   0.000731 &     0.0270\\
2$_{12}$--1$_{01}$  $J$=2--1 & 107.8573 (2) &   0.0184 &     1.13\\
2$_{12}$--1$_{01}$  $J$=2--2 & 107.4832 (2) &   0.00603 &     0.367\\
{\bf 2$_{{\bf 12}}$--1$_{{\bf 01}}$  ${ J}$=3--2} & {\bf 107.7203 (2)} &   {\bf 0.0244} &     {\bf 2.10}\\
\hline
\end{tabular} 
\end{center}
\end{table}

Frequencies for low-lying rotational transitions have been calculated from Laser Magnetic Resonance (LMR) laboratory spectra by \citet{sears} with a quoted accuracy of 5~MHz (0.6 km~s$^{-1}$). In order to confirm the lines and to calculate accurate line strengths we have used new measurements of the sub-mm lines \citep{michael, bruenken} to determine more accurate frequency values. Table~\ref{freqs} shows the calculated wavelengths, Einstein coefficients and line strengths (averaging over the hyper-fine structure in ortho transitions). The accuracy of the calculated wavelengths is highest for those transitions actually measured in the lab \citep[2$_{11}$--2$_{02}$ (153~$\mu$m) and 1$_{10}$--1$_{01}$ (156~$\mu$m); ][]{bruenken}. The uncertainty in the calculated Einstein values and line strengths is dominated by that of the dipole moment. We have followed previous authors and used a value of 0.57~D \citep[derived from an ab initio calculation by][]{bunker}. The calculation strongly depends on the electronic and geometric structure of the molecule and inserting the most recent values for the geometry suggests that the uncertainty must be at least 0.02~D (leading to $\sim$5\% uncertainty in the Einstein coefficients). The line strength in Table~\ref{freqs} is defined in the same way as by \citet{sears}, allowing a direct comparison. This shows that the strengths calculated here are half those previously published. However, our results can reproduce the values for the 1$_{11}$--2$_{02}$ (317~$\mu$m) transition given by \citet{michael} and agree with the calculations of \citet{chandra} (both of these authors used an independent method to calculate their line strengths). This indicates that the line strengths quoted in the paper by \citet{sears} are likely to be in error by a factor of two (T. Sears, private communication).

\section{Observations and data reduction}

The \object{Sgr~B2} observations were carried out as part of a wide spectral survey using the ISO LWS Fabry-P\'{e}rot (FP) mode, L03. The whole LWS spectral range, from 47 to 196~$\mu$m (6.38--1.53~THz), was covered using 36 separate observations with a spectral resolution of 30--40~km~s$^{-1}$. The first results from this survey have recently been presented by \citet{ceccarelli,polehampton_b,vastel_b,polehampton_c}. We also analysed additional L04/L03 mode observations towards \object{Sgr~B2}, \object{W49~N}, \object{Sgr~A$^{\ast}$} and \object{NGC7023} and one grating L01 observation towards \object{W49~N}, downloaded from the ISO Data Archive\footnote{see http://www.iso.vilspa.esa.es/ida.}. Details of the ISO TDT numbers of all the observations used in this paper are shown in Table~\ref{tdts}.

\begin{table}[!t]
\caption{\label{sourcepos}Coordinates of observed sources.}
\leavevmode \footnotesize
\begin{center}
\begin{tabular}[!t]{ccc}
\hline
\hline
Source                   & RA (J2000)                                         & Dec. (J2000) \\
\hline
\object{Sgr~B2}          & $17^{\mathrm{h}}47^{\mathrm{m}}21.75^{\mathrm{s}}$ & $-28\degr 23\arcmin 14.1\arcsec$ \\
\object{W49~N}           & $19^{\mathrm{h}}10^{\mathrm{m}}13.55^{\mathrm{s}}$ & $+09\degr 06\arcmin 14.7\arcsec$ \\
\object{Sgr A$^{\ast}$}  & $17^{\mathrm{h}}45^{\mathrm{m}}39.97^{\mathrm{s}}$ & $-29\degr 00\arcmin 23.6\arcsec$ \\
\object{NGC7023}         & $21^{\mathrm{h}}01^{\mathrm{m}}36.91^{\mathrm{s}}$ & $+68\degr 09\arcmin 48.2\arcsec$ \\
\hline
\end{tabular}
\end{center}
\end{table}

The LWS beam has an effective diameter of $\sim80\arcsec$ \citep{gry} and was centred on the coordinates shown in Table~\ref{sourcepos}. The coordinates chosen for \object{Sgr~B2} were slightly offset from the main FIR peak at \object{Sgr~B2~(M)} to ensure that \object{Sgr~B2~(N)} was excluded from the beam. Three positions were observed towards NGC7023 - the illuminating star (coordinates given in Table~\ref{sourcepos}), the NW PDR ($-24.5\arcsec$, $+39.2\arcsec$) and the SW PDR ($-25.8\arcsec$, $-63.1\arcsec$).

The L03 observations towards \object{Sgr~B2} were reduced using the ISO Offline pipeline (OLP) version 8 with the remaining observations processed with OLP version 10 (for FP observations there is no significant difference between OLP 8 and 10). Further processing was then applied interactively using routines that appeared as part of the LWS Interactive Analysis package version 10 \citep[LIA10:][]{lim_d} and the ISO Spectral Analysis Package \citep[ISAP:][]{sturm}. The interactive processing included the calculation of accurate dark currents (including stray light), careful removal of the LWS grating profile from the spectra and removal of glitches caused by cosmic ray impacts \citep[see ][]{polehampton_b,polehampton_c}. The wavelength scale of each observation was then corrected to the local standard of rest and data co-added for each line (small multiplicative factors were sometimes necessary to match each observation in flux before co-adding). The final uncertainty in wavelength is less than 0.004~$\mu$m (or 11~km~s$^{-1}$) corresponding to the error in absolute wavelength calibration \citep[see ][]{gry}. Finally, a polynomial baseline was divided into the data to obtain the line-to-continuum ratio. For the \object{Sgr~B2} data, a 3rd order polynomial was necessary to fit the continuum around and between the detected lines. For the L04 data towards \object{W49~N} where the continuum coverage is much lower, a 1st order baseline was used. The use of low order polynomials and careful masking of the detected lines for the fit ensured that no spurious features were introduced into the data by the division. This step in the reduction is very useful because it effectively bypasses the large uncertainty in absolute flux level caused by multiplicative calibration steps \citep[see][]{swinyard_b}. The remaining error is dominated by statistical noise in the data with a small additional uncertainty due to the dark current determination and continuum fit.

In Sect.~\ref{othersources}, we use one grating L01 observation towards \object{W49~N} to determine the column density of CH. This observation was reduced using OLP 10, and corrected for saturation in the LWS detectors (necessary due to the strength of the \object{W49~N} thermal continuum) by T. Grundy at the UK ISO Data Centre. The dataset was further reduced using the LWS L01 post-processing pipeline which performs several additional corrections detailed in \citet{lloyd_d}. This allowed us to calculate an accurate value for the line-to-continuum ratio in the line.

\section{\object{Sgr~B2} results}

We searched the \object{Sgr~B2} spectrum for all the low-lying rotational lines of both ortho- and para-CH$_{2}$ occurring within the survey range (47--196~$\mu$m). Figure~\ref{CHfit} shows the data around the three lowest energy transitions (fits to the line shapes are also shown - these are described in Section~\ref{ch2colden}). At the spectral resolution of the LWS FP, the fine structure splitting of each rotational transition is resolved but the spacing of the hyperfine components in ortho-CH$_{2}$ ($<10$~km~s$^{-1}$) is too small to be separated. When analysing ortho transitions we have used wavelengths averaged over the hyperfine structure weighted by $A_{ij}$. The wavelengths used are shown in Table~\ref{freqs}.

The survey also shows strong absorption due to CH (shown in the top panel of Fig.~\ref{CHfit}). We have used these lines as a basis for fitting the observed CH$_{2}$ line shapes to fix the relative contribution from features along the line of sight. This was necessary because the signal-to-noise in the CH$_{2}$ detections is not high enough to allow a full fit accounting for different line of sight components separately. The method assumes that there is a constant [CH/CH$_{2}$] ratio applicable to all line of sight components.

\subsection{CH}

Two prominent absorption features due to CH are present in the spectrum due to its $^{2}\Pi_{1/2}$ $J$=3/2--1/2 transition from the ground state to the first rotational level, $\sim$96~K above ground \citep[see also ][]{cernicharo_b,goicoechea_d}. These are due to the $\Lambda$-doublet type splitting of each rotational level and occur at wavelengths of 149.09~$\mu$m and 149.39~$\mu$m \citep{davidson}. The two detected lines are shown in the top panel of Fig.~\ref{CHfit} and are broad with absorption in the range $-150$ to $+100$~km~s$^{-1}$. This is due to galactic spiral arm clouds between the Sun and Galactic Centre \citep[centred at velocities $-100$ to $+30$~km~s$^{-1}$, e.g. see][]{greaves94} as well as to \object{Sgr B2} itself. The peak absorption occurs near the velocity of \object{Sgr~B2} at $\sim+65$~km~s$^{-1}$. These lines have previously been detected with the Kuiper Airborne Observatory (KAO) by \citet{stacey} but the ISO data provide a significant improvement in the signal-to-noise ratio as well as complete coverage of the neighbouring continuum. No higher transitions of CH were detected in the survey above the noise.

In order to determine the CH column density for each velocity component in the line of sight towards \object{Sgr~B2}, a high resolution model of the line shape was constructed, convolved to the LWS resolution and adjusted to obtain a best fit. This was carried out in the same way as used for OH lines \citep[see][Polehampton et al. in preparation]{polehampton_c}. \ion{H}{i} 21~cm absorption measurements \citep{garwood} were used to fix the velocity and line width of 10 line of sight components and then optical depths were adjusted in a multi-parameter fit to find the minimum in $\chi^{2}$. The optical depth for each component, $\tau$,  was calculated from the line-to-continuum ratio using
\begin{equation}
\label{eqn_abs}
I = I_c \exp{(-\tau)}
\end{equation}
where $I_c$ is the intensity of the continuum. This method assumes that the same velocity components seen in the \ion{H}{i} spectrum are present in CH and CH$_{2}$ \citep[but the fit does not depend on the \ion{H}{i} optical depths derived by][]{garwood}. This is likely because the atomic material is seen to be associated with molecular gas at the same velocities \citep[e.g.][]{vastel_b}. Each fitted component probably still represents a mean over many narrower features such as those seen in CS absorption with velocity widths $\sim$1~km~s$^{-1}$ \citep{greaves94}. The final fit is shown in the top panel of Fig.~\ref{CHfit}.

Column densities for each velocity component are shown in Table~\ref{CHcolden}. These were calculated assuming a Doppler line profile with Maxwellian velocity distribution \citep[e.g.][]{spitzer},
\begin{equation}
\label{eqn_colden}
N_j= \frac{8\pi\sqrt{\pi}}{2\sqrt{\ln2}}~10^{17}~\frac{\tau_{0}~\Delta{v}} {A_{i\,j}~\lambda_{i\,j}^{3}~g_{i}/g_{j}}
\end{equation}
where $N_j$ is the column density in the lower level, $\tau_{0}$ is the optical depth at line centre, $\Delta{v}$ is the line width in km~s$^{-1}$, $A_{ij}$ is the Einstein coefficient for spontaneous emission, $\lambda_{ij}$ is the wavelength in $\mu$m and $g_{i}$ is the statistical weight of state $i$.

\begin{table}[!t]
\caption{\label{CHcolden} Results of our fit of the CH lines towards \object{Sgr~B2} are given in column (A). The velocities and line widths are taken from the \ion{H}{i} data of \citet{garwood}. Column (B) gives the previous estimates of CH column densities from the KAO measurements of \citet{stacey} and columns (C) and (D) give the results from the radio $\Lambda$-doublet lines from \citet{genzel} and \citet{andrew}.}
\leavevmode \footnotesize
\begin{center}
\begin{tabular}[!t]{cccccc}
\hline
\hline
LSR Velocity    &   FWHM        & \multicolumn{4}{c}{$N(\mathrm{CH})$ (10$^{14}$~cm$^{-2}$)} \\
(km~s$^{-1}$)   & (km~s$^{-1}$) &  (A)         & (B)  & (C) & (D) \\
\hline
$-107.6$        & 7     & 0.9$\pm$0.7  & 4.2  &     &      \\
$-81.7$         & 28    & 2.6$\pm$1.1  &      &     & 2.5  \\
$-51.9$         & 17    & 0.3$\pm$0.15 &      &     &      \\
$-44.0$         & 8     & 1.5$\pm$0.6  & 3.6  &     & 1.3  \\
$-24.4$         & 14    & 1.9$\pm$0.4  &      &     & 0.3  \\
$+1.1^a$        & 19    & 2.0$\pm$0.3  & 3.2  &     & 1.1  \\
$+15.7 $        & 7     & 3.1$\pm$0.8  &      &     & 1.5  \\
$+31.4 $        & 21    & 1.5$\pm$1.0  &      &     &      \\
$+52.8/+66.7^b$ & 11/16 & 9.3$\pm$0.9  & 5.2  & 3.1 & 9.4$^c$  \\
\hline
Total           &       & 20.1         & 15.2 &     & 16.1  \\
\hline
\end{tabular}
\end{center}
$^a$ This velocity is corrected from an error in \citet{garwood} where it is given as $+11$~km~s$^{-1}$\\
$^b$ The \ion{H}{i} data resolve 2 components in \object{Sgr~B2}. However, our fit requires
only one of these (at 66.7~km~s$^{-1}$) to reproduce the ISO spectrum.\\
$^c$ This is the combined value for all components $>50$~km~s$^{-1}$ assuming the emission fills the beam,
as described in the text by \citet{andrew}.
\end{table}

No higher transitions of CH have been observed in the spectrum even at the velocity of \object{Sgr~B2} itself, showing that the level populations are characterised by a low rotation temperature (i.e. sub-thermal excitation) and the ground state population is a good measure of the total column density. Using a 2$\sigma$ detection limit for the $^{2}\Pi_{1/2}$ $J$=5/2--3/2 CH transition (116~$\mu$m), we calculate $T_{\mathrm{rot}}<20$~K. The fitted optical depths show that the absorption is generally optically thin, although in the \object{Sgr~B2} component at 67~km~s$^{-1}$ it is marginally optically thick ($\tau$=2.4). The results of the fit are shown in Fig.~\ref{CHfit} and Table~\ref{CHcolden}, with errors determined by examining $\chi^{2}$ as pairs of optical depths were varied about their best fit values.

One way to check the opacity of \object{Sgr~B2} is to use the less abundant $^{13}$CH isotopomer whose $^{2}\Pi_{1/2}$ $J$=3/2--1/2 transition has components at 149.79~$\mu$m and 150.09~$\mu$m \citep{davidson2}. \citet{langer2} find a $^{12}$C/$^{13}$C ratio towards \object{Sgr~B2} of $24\pm1$ using observations of $^{12}$C$^{18}$O and $^{13}$C$^{18}$O. However, the $^{13}$CH  lines are not detected in the ISO spectrum to a limit of 1.3\% of the continuum (2$\sigma$ level). This leads to $^{12}$CH/$^{13}\mathrm{CH}>36$ at the velocity of \object{Sgr~B2}. This is higher than the value derived from CO but this could be due to isotopic fractionation which is expected to increase the ratio from the true $^{12}$C/$^{13}$C value \citep[see][]{langer}. In the following analysis, we use the column density for \object{Sgr~B2} without adjustment.

The final CH column densities in Table~\ref{CHcolden} compare well with previous values derived from the KAO observations of same FIR lines by \citet{stacey}. \citet{goicoechea_d} have also examined ISO data of CH and found absorption extended across the whole surrounding region. They calculate total column densities in the range (0.8--1.8)$\times$10$^{15}$~cm$^{-2}$, peaking at the central \object{Sgr B2 (M)} and \object{Sgr B2 (N)} positions. Radio observations of the CH $\Lambda$-doublet lines by \citet{andrew} are in very good agreement with the individual velocity features that we have fitted - the small discrepancies may be due to their much larger beam size (8.2$\arcmin$), which also included the (N) position. The radio lines were also observed by \citet{genzel}, however, they only give the column density for \object{Sgr~B2} itself at $+65$~km~s$^{-1}$ (although they also observed emission from the other velocity components). They derived the column density from the 3264~MHz line giving a value three times lower than ours (however, their weaker data for the 3335~MHz line would give a higher column density).

\subsection{Ortho-CH$_{2}$}

The lowest ortho level of CH$_{2}$ occurs at the ground level, $N$=0. We clearly detect absorption from the 1$_{11}$--0$_{00}$ $J$=2--1 and $J$=1--1 fine structure components at 127~$\mu$m (see Fig.~\ref{CHfit}). The third component, $J$=1--0, is not detected above the noise in the spectrum. The observed features have a similar shape to the CH lines, indicating the presence of absorption along the whole line of sight.

The next highest energy level occurs $\sim$67~K above ground. There are two transitions from this level that occur within the ISO wavelength range at 153~$\mu$m (2$_{11}$--2$_{02}$) and 94~$\mu$m (3$_{13}$--2$_{02}$), and one that occurs outside the range at 317~$\mu$m (1$_{11}$--2$_{02}$). The strongest fine-structure component of these is the 2$_{11}$--2$_{02}$ $J$=3--3 line at 153.353~$\mu$m. However, this wavelength is very close to the $N$=2--1 $J$=3--2 transition of NH at 153.348~$\mu$m (separation $\sim$10~km~s$^{-1}$) and there is a strong absorption line detected \citep[see][]{cernicharo_d}. The identification of this feature with NH is secure because we also detect its $J$=2--1 and $J$=1--0 lines with relative depths that agree extremely well with the predicted line strengths \citep[measured values are 1.0/0.55/0.27 compared to the predicted relative line strengths of 1.0/0.54/0.23 from the  JPL line catalogue;][]{pickett}. Furthermore, the next strongest CH$_{2}$ transition at 153.814~$\mu$m is not detected. This shows that the major contribution to the 153.35~$\mu$m line must be from NH rather than CH$_{2}$.

The next strongest transition is 3$_{13}$--2$_{02}$ and the $J$=4--3 line (93.662~$\mu$m) should be almost as strong as the 153.35~$\mu$m line discussed above. There appear to be some features at roughly the correct velocity at a level $\sim3\sigma$ above the statistical noise (see Fig.~\ref{CHfit} but note that the error bars shown include systematic uncertainty). 

No higher level ortho transitions were detected above the noise in the data.

\subsection{Para-CH$_{2}$ \label{parach2}}

\begin{figure}[!t]
\includegraphics[height=17.5cm]{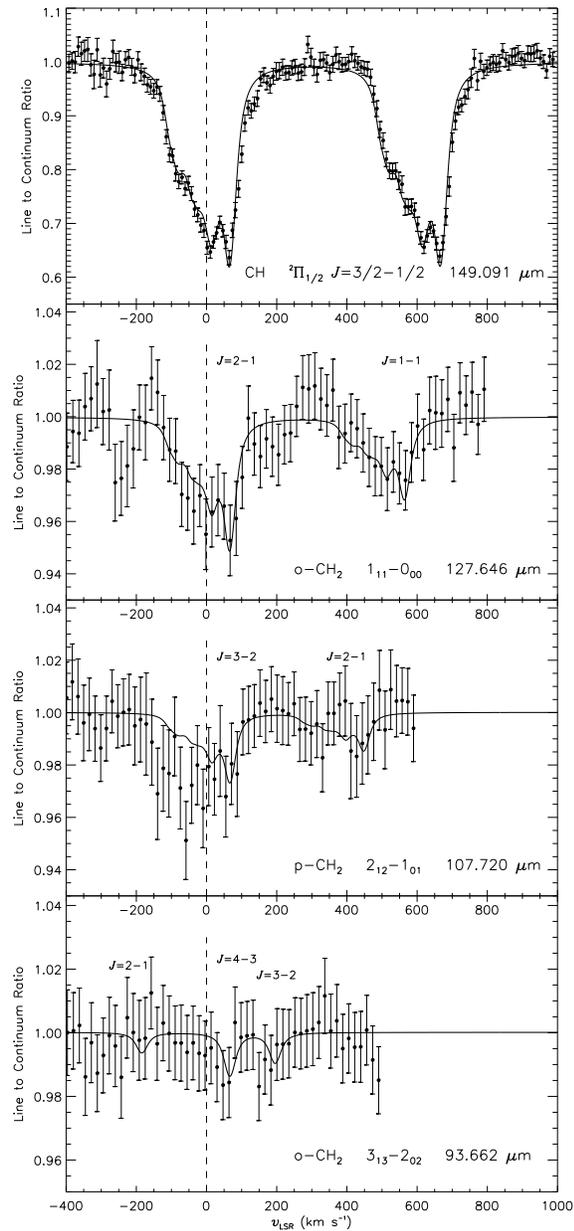}
\caption{\label{CHfit}
{\bf Upper panel:} Fit of the CH line towards \object{Sgr~B2} with the data binned at a quarter of the resolution element. 
{\bf Lower panels:} In the lower three panels, a model based on the CH line shape has been fitted to the data around the three lowest energy transitions of CH$_{2}$. The relative contribution of each velocity component in these fits was fixed to the CH values in Table~\ref{CHcolden}, except for the 94~$\mu$m (3$_{13}$--2$_{02}$) transition in the bottom plot which only includes the velocity component due to \object{Sgr B2} itself (see text). The fit was carried out to the strongest fine structure component in each case. Other (sometimes undetected) components are included in the model using their expected relative line strengths from Table~\ref{freqs}. The data are binned at half of the resolution element and the errors reflect both statistical noise and systematic error due to the dark current subtraction and continuum fit. The extra absorption feature seen at $-250$~km~s$^{-1}$ relative to the 127~$\mu$m (1$_{11}$--0$_{00}$) transition appears to be real but is not identified.}
\end{figure}

The lowest energy state for para-CH$_{2}$ occurs at $\sim$23~K above ground (see Fig.~\ref{levels}) and has two transitions at 156~$\mu$m (1$_{01}$--1$_{10}$) and 107~$\mu$m (2$_{12}$--1$_{10}$). The strongest fine-structure component of these is the 2$_{12}$--1$_{10}$ $J$=3--2 transition at 107.720~$\mu$m. There is an absorption feature at approximately the correct wavelength for this line, although it appears to be wider than expected in its short wavelength wing. The observed absorption depth is also consistent with the data at the position of the next strongest component, $J$=2--1, at 107.857~$\mu$m (see Fig.~\ref{CHfit}). Using this detection to predict the optical depth of the strongest 1$_{01}$--1$_{10}$ transition at 156.332~$\mu$m ($J$=2--2) gives a value consistent with the noise level in the spectrum.

No higher para transitions were detected above the noise in the data.

\subsection{CH$_2$ column densities \label{ch2colden}}

In order to fit the CH$_{2}$ lines, we assumed that the relative column densities between each velocity feature were the same as for CH. This fixes the shape of the line and allows a single parameter fit to determine an 'average' CH/CH$_{2}$ column density ratio for each CH$_{2}$ energy level. The fit was carried out to the strongest fine structure component in each case, with the relative importance of the other (sometimes undetected) fine structure transitions fixed using the line strengths from Table~\ref{freqs}. The results of these fits are shown in the lower 3 panels of Fig.~\ref{CHfit} (including the undetected transitions, to show that their predicted absorption is consistent with non-detection in the data).

This method worked well for the strongest transition (1$_{11}$--0$_{00}$) at 127~$\mu$m, where both the line shape and relative strength of the $J$=2--1 and $J$=1--1 components show a good fit. A detection limit for the $J$=0--1 component is also consistent with the expected line strengths of 1.0/0.6/0.2 for $J$=2--1, $J$=1--1 and $J$=0--1 respectively. The fact that the line shape agrees well with the data indicates that the [CH/CH$_{2}$] ratio does not vary significantly between the different line of sight components.

The para transition at 107.7~$\mu$m (2$_{12}$--1$_{01}$) does not show such good agreement in line shape with the model. As mentioned in Sect.~\ref{parach2}, the main line appears wider than expected, although the noise is high in this part of the spectrum. It is unlikely that this extra absorption is associated with the CH$_{2}$ line as this would mean that the negative velocity features would have to be stronger than the component due to \object{Sgr~B2} itself and this is not observed in other lines. It could be due to overlap with an unidentified feature (although we have not been able to find a likely candidate) or an instrumental effect \citep[e.g. the overlap of mini-scans can sometimes cause problems for wide lines - see][]{polehampton_c}. A final possibility is that the extra absorption could have been introduced when dividing by the polynomial fit to the continuum. However, we minimised this effect by using a low order polynomial (3rd order) to carefully fit only the continuum around features that already appeared in the raw data. Any variation in the fitted baseline is on a larger scale than the excess absorption in the line. In fitting the line, we have only considered velocities above 0~km~s$^{-1}$. The actual line shape is not well constrained by these data and our fitting method of fixing the relative contribution of each velocity to that of CH may overestimate the absorption due to the line of sight clouds.

In order to fit the data around the transition at 94~$\mu$m (3$_{13}$--2$_{02}$), we modified the line template to include only the velocity component at $+67$~km~s$^{-1}$ from Table~\ref{CHcolden}. This is because we only expect a significant population in the higher levels at the velocity of \object{Sgr B2} itself. In \object{Sgr~B2} itself, excitation to the 2$_{02}$ level ($\sim$67~K above ground) can be provided by FIR photons from the strong dust continuum, whereas in the line of sight clouds there is no strong radiation field and the densities are too low for collisional excitation to be important. The bottom panel in Fig.~\ref{CHfit} shows the results of the fit.

The small discrepancy in velocity between fit and data can be explained by the fact that the main feature is only 3$\sigma$ above the noise. The uncertainty in the exact line wavelength is also relatively high for this transition as it has never been measured in the lab either in Cologne or by \citet{sears}. The difference between the value quoted in Table~\ref{freqs} and that by \citet{sears} is $\sim$3~km~s$^{-1}$.

The final column densities for the lowest ortho and para levels of CH$_{2}$ were determined directly from the fitted CH/CH$_{2}$ ratio for each level, assuming that the distribution over fine structure levels is determined by the line strengths given in Table~\ref{freqs}. The final best fit values summed over all velocities are, 
\begin{center}
$N(0_{00}) = (2.9\pm0.3)\times10^{14}$~cm$^{-2}$\\
$N(1_{01}) = (3.4\pm0.9)\times10^{14}$~cm$^{-2}$
\end{center}

The column density in the $2_{02}$ level at the velocity of \object{Sgr~B2} is, $N(2_{02})_{\rm{sgrB2}} = (1.2\pm0.5)\times10^{14}$~cm$^{-2}$. Treating this as an upper limit allows us to make a comparison with the ortho ground state (using only the \object{Sgr~B2} component) and estimate the excitation temperature for \object{Sgr~B2}, giving $T_{\rm{rot}}<$40$^{+14}_{-11}$~K. In order for absorption to be seen, this excitation temperature must be lower than the temperature of the dust producing the FIR background. In \object{Sgr~B2} the dust temperature has been found to be $\sim30$~K \citep{goicoechea_c}, and so our estimated excitation temperature is consistent within its errors. At this temperature we do not expect any higher levels to be significantly populated and do not detect any of the other higher transitions that occur within the spectral survey range (2$_{20}$--1$_{11}$ at 50.5~$\mu$m; 2$_{21}$--1$_{10}$ at 50.8~$\mu$m and 3$_{12}$--3$_{03}$ at 148.8.5~$\mu$m).

Assuming that only the first three energy levels in \object{Sgr~B2} are populated, the ortho-to-para ratio in the $+67$~km~s$^{-1}$ component is $1.6^{+0.9}_{-0.6}$. The equilibrium value of the ratio for CH$_{2}$ should be $\geq$3 for low temperatures. This difference could be a relic of the formation process with the current ortho-to-para ratio fixed at formation and/or set by the ratio of the parent species. Alternatively, it could indicate that not all of the 2$_{12}$--1$_{01}$ absorption seen at 107.7~$\mu$m is associated with CH$_{2}$, possibly due to blending with another line. Reducing the para column density by approximately a factor of 2 would lead to an ortho-to-para ratio of 3.

Overall, the column density summed over all velocities in both ortho and para states is $N(\rm{CH_{2}})=(7.5\pm1.1)\times10^{14}$~cm$^{-2}$, giving a final ratio with CH of [CH/CH$_{2}$] = 2.7$\pm$0.5. If the para column density is reduced to make the ortho-to-para ratio equal to 3, the [CH/CH$_{2}$] ratio would be increased to 3.7.

\section{Abundances}

To discuss CH and CH$_2$ abundances, we need an estimate of the hydrogen column density towards \object{Sgr~B2}.  

In the Galactic spiral arm clouds along the line of sight ($-$100 to $+$30~km~s$^{-1}$), HCN and CS measurements suggest average densities of 200~cm$^{-3}$, but also the presence of gas up to 10$^{4}$~cm$^{-3}$ \citep{greaves95}. Their structure is probably similar to photodissociation regions (PDRs) with a molecular core and atomic skin at the surface, and UV illumination provided by the mean interstellar radiation field \citep{vastel_b}.  $N({\rm H}_2)$ has been estimated to be 5, 9, 4 and $14\times10^{21}$~cm$^{-2}$ respectively in the $-100$, $-40$, $-25$ and 0~km~s$^{-1}$ features \citep{greaves96}, with corresponding atomic hydrogen column densities approximately equal to 2, 4, 2 and $8\times10^{21}$~cm$^{-2}$ \citep{vastel_b}. In order to compare with diffuse cloud models, we calculate the abundances taking into account the total hydrogen particle column, $N_{\rm{H}}=N(\rm{H})+2N(\rm{H_2})$, although it is difficult to determine whether the observed CH and CH$_{2}$ co-exists with both atomic and molecular hydrogen regions - they are probably confined to one or more specific layers within each cloud. Using our fitted column densities for CH in Table~\ref{CHcolden}, we calculate an approximate abundance of $N({\rm CH})/N_{\rm H}\sim$(0.6--3)$\times$10$^{-8}$ for these clouds. According to our average [CH/CH$_{2}$] ratio, $N({\rm CH}_2)/N_{\rm H}$ is thus (0.2--1.1)$\times$10$^{-8}$.

The situation is more complicated at the velocity of \object{Sgr~B2} as there may be contributions to the absorption from layers with widely different conditions. \citet{comito} have found that a significant fraction of the water absorption towards \object{Sgr~B2} must be due to a hot (500--700~K), low density layer \citep[also observed in ammonia lines; ][]{ceccarelli, huttemeister_b} but that there must also be absorption from the warm (40--80~K) envelope. \citet{goicoechea_b} have derived a large OH abundance at the velocity of \object{Sgr~B2} suggesting that there is a strong UV field producing clumpy PDRs and a temperature gradient from 40--600~K through the envelope. The total H$_{2}$ column density in front of the FIR continuum has been estimated from the FIR spectrum to be (2--10)~$\times$10$^{23}$~cm$^{-2}$ \citep{goicoechea_c}. Using this to calculate approximate abundances gives $N({\rm CH})/N_{\rm H}\sim$(0.5--2)$\times10^{-9}$ and $N({\rm CH}_2)/N_{\rm H}\sim$(0.2--0.7)$\times$10$^{-9}$.

\section{CH$_{2}$ in other sources \label{othersources}}

We have also searched the ISO Data Archive for observations towards other sources that may show the low-lying CH$_{2}$ lines.

\subsection{\object{W49~N}}

Observations that cover the wavelengths of the CH$_{2}$ lines were carried out using the LWS L04 mode towards the active star forming region, \object{W49~N}. This is the strongest IR peak in the \object{W49~A} molecular cloud complex which is located at 11.4~kpc from the Sun and 8.1~kpc from the Galactic Centre \citep{gwinn}. Its spectrum shows a strong thermal continuum in the FIR with a peak near 60~$\mu$m  \citep{vastel_c}. Line emission associated with the molecular cloud itself is centred at 8~km~s$^{-1}$ \citep[e.g.][]{jaffe}, but there are also several features due to intervening gas at velocities between 16 and 75~km~s$^{-1}$ \citep[e.g.][]{nyman}. \citet{vastel_a} have mapped the CO emission in the ISO LWS beam and found 7 velocity components in the range 35--70~km~s$^{-1}$. Recently, \citet{plume} have measured H$_{2}$O towards \object{W49}  and the same features show up in absorption. They arise because the line of sight intersects the Sagittarius spiral arm in two places \citep[e.g.][]{greaves94}.

\begin{figure}[!t]
\resizebox{\hsize}{!}{\includegraphics{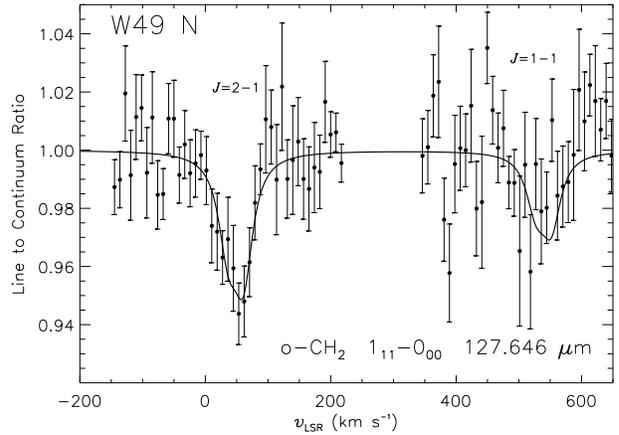}}
\caption{\label{w49}Data around the $J$=2--1 (127.6~$\mu$m) and $J$=1--1 (127.8~$\mu$m) fine structure components of the $1_{11}$--$0_{00}$ CH$_{2}$ transition toward \object{W49~N}. The solid line shows a model derived from a fit to the $J$=2--1 component and the expected relative line strength of the $J$=1--1 component.}
\end{figure}

The strongest ortho-CH$_{2}$ 1$_{11}$--0$_{00}$ fine structure component at 127.6~$\mu$m ($J$=2--1) is clearly detected in absorption in the LWS FP spectrum. However, the spectral resolution and signal-to-noise ratio in the data do not allow a detailed fit using all the velocity components observed in CO and H$_{2}$O lines. Therefore, we performed a simplified fit using two components to represent the strongest features seen in the H$_{2}$O spectrum of \citet{plume}. We fixed the velocities (and FWHM) to be 37~km~s$^{-1}$ (10~km~s$^{-1}$) and 61~km~s$^{-1}$ (7~km~s$^{-1}$) and found the best fitting optical depths after convolving to the LWS resolution. Lower velocity components were not necessary to reproduce the line shape, showing that all the absorption comes from the line of sight clouds. Only the $J$=2--1 component was used for the fit. The result is shown in Fig.~\ref{w49}, which also shows a prediction for the $J$=1--1 (127.858~$\mu$m) fine structure component based its expected relative line strength. Column densities for the 0$_{00}$ level, calculated from the best fitting optical depths and equation~\ref{eqn_colden}, are $(0.8\pm0.4)\times10^{14}$~cm$^{-2}$ and $(1.2\pm0.4)\times10^{14}$~cm$^{-2}$ at 37~km~s$^{-1}$ and 61~km~s$^{-1}$ respectively.

The para transition at 107~$\mu$m (2$_{12}$--1$_{01}$) is not detected above the noise. Assuming that only the two lowest levels are populated and using a 2$\sigma$ limit on the line depth, gives an ortho-to-para ratio $>$2.1. 

The CH line at 149~$\mu$m was not observed toward \object{W49~N} using the FP mode of the LWS, but it was measured with the lower resolution grating mode. At this resolution the $\Lambda$-doublet line splitting, as well as the line of sight structure, is not resolved and we can only estimate a total CH column density that may include absorption associated with \object{W49} itself, giving $8.4\times10^{14}$~cm$^{-2}$. \citet{sume} have observed CH radio $\Lambda$-doublet lines towards \object{W49~A} and they find column densities of $8\times10^{13}$cm$^{-2}$ and $10\times10^{13}$cm$^{-2}$ in the velocity intervals 30--45~km~s$^{-1}$ and 54--72~km~s$^{-1}$. \citet{rydbeck} also observed the $\Lambda$-doublet lines, giving column densities of $7\times10^{13}$cm$^{-2}$ and $18\times10^{13}$cm$^{-2}$. Both authors show that these two velocity intervals give the most important CH contributions. This means the extra column density seen in the LWS grating measurement probably also has highest contribution from these ranges and the radio measurements may be an underestimate (probably due to the much larger beam size of 15$\arcmin$ used for these observations).

Using the average of the two radio CH measurements and combining the ortho column density and para upper limit for CH$_{2}$, gives a lower limit for the [CH/CH$_{2}$] ratio of 1.4 and 0.8 in the two velocity intervals.

\begin{figure}[!t]
\resizebox{\hsize}{!}{\includegraphics{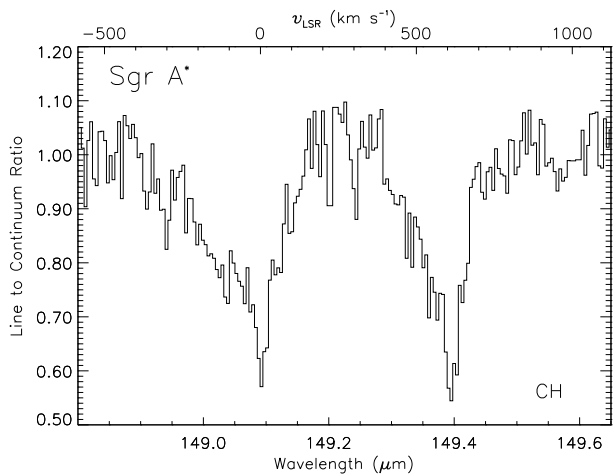}}
\caption{\label{sgra}The CH $^{2}\Pi_{1/2}$ $J$=3/2--1/2 transition observed towards \object{Sgr~A$^{\ast}$} showing two resolved $\Lambda$-doublet components.}
\end{figure}

\subsection{\object{Sgr~A$^{\ast}$}}

Observations using the LWS FP mode were also made towards \object{Sgr~A$^{\ast}$} at the Galactic Centre. It has a FIR continuum showing strong thermal emission from dust, peaking at $\sim$50~$\mu$m. Observations of several hydrocarbons have been reported towards this source, including CH$_{3}$ \citep{feuchtgruber}. We have searched LWS FP observations for the lines of CH$_{2}$ and CH and Fig.~\ref{sgra} shows the observed absorption due to the $^{2}\Pi_{1/2}$ $J$=3/2--1/2 (149~$\mu$m) rotational transition of CH. These lines show peak absorption near 0~km~s$^{-1}$, probably due to local gas. There is also absorption at negative velocities due to galactic spiral arm clouds. This picture is consistent with observations of CH $\Lambda$-doublet lines towards \object{Sgr A}, which showed that the strong Galactic Centre component normally observed in other molecules at $+50$~km~s$^{-1}$ is relatively weak in CH \citep{genzel}. At the wavelength of the ortho CH$_{2}$ 1$_{11}$--0$_{00}$ transition (127~$\mu$m), the signal-to-noise ratio in the data is rather low compared to the \object{Sgr~B2} observations and we do not detect any absorption. The para 2$_{12}$--1$_{01}$ transition at 107.7~$\mu$m is also not detected. Combining 2$\sigma$ upper limits for both transitions, we calculate a lower limit to the [CH/CH$_{2}$] ratio at 0~km~s$^{-1}$ of 2.6.

\begin{table*}[!t]
\caption{\label{comp} Comparison of model predictions for diffuse and dense clouds with the observed abundances (compared to $2N(\rm{H_{2}})+N(\rm{H})$) of CH and CH$_{2}$ ($X$(CH) and $X$(CH$_{2}$)). The diffuse cloud model is for the line of sight to $\zeta$~Per \citep{vandishoeck86}: vDB86. The quoted density and temperature are for the centre of the cloud in the model. The dense cloud models are steady state values corresponding to the gas phase ``new standard model'' of \citet{lee}: L96 and the gas-grain model of \citet{ruffle}: RH01 including photochemistry (P1/A at 10$^{8}$~yr). The PDR model is for a density of 10$^{5}$~cm$^{-3}$ and radiation field $\sim$650--900 times the average interstellar value \citep{jansen95}: J95. The observational results derived from our ISO measurements are shown as well as previous results from \citet{lyu} and \citet{hollis}.}
\leavevmode \footnotesize
\begin{center}
\begin{tabular}[!t]{lcccccc}
\hline
\hline
                 &                     & Density     & Temperature &  $X$(CH)    & $X$(CH$_{2}$) & $[{\rm CH}/{\rm CH}_2]$ \\
                 &                     &(cm$^{-3}$)  & (K)         & ($10^{-8}$) & ($10^{-8}$)   &  \\
 \hline
Model        & Diffuse (vDB86)         & 500             & 25 & 1.6       & 2.4        & 0.67 \\ 
predictions: & Dense    (L96)          & $10^3$          & 10 & 0.25      & 1.6        & 0.018  \\ 
             & Dense      (L96)        & $10^5$          & 10 & 0.0007    & 0.039      & 0.16  \\ 
             & Dense   (RH01)          & $2\times10^{4}$ & 10 & 0.1       & 1.2        & 0.08 \\ 
             & PDR    (J95)            & 10$^5$          &    & 1.4       & 0.7        & 2.0  \\ 
 \hline
Results from & Sgr B2 Line of Sight           &            &         & 0.6--3    & 0.2--1.1   & 2.7$\pm$0.5   \\
this work:   & \object{Sgr~B2}                &                     &         & 0.05--0.2 & 0.02--0.07 & 2.7$\pm$0.5   \\ 
             & \object{W49~N} 37~km~s$^{-1}$           &            &         &           &            & $>$1.4 \\  
             & \object{W49~N} 61~km~s$^{-1}$           &            &         &           &            & $>$0.8\\
             & \object{Sgr A$^{\ast}$}                 &            &         &           &            & $>$2.6\\  
 \hline
Previous     & $\zeta$~Oph                    &            &          & 1.8$^{b}$   & 4.3$^{a}$  & 0.4$^{b}$ \\ 
observational& HD154368                       &            &          & 1.9$^{d}$   & 0.79$^{c}$ & 2.4$^{d}$ \\ 
results:     & Orion-KL                       & $\sim10^5$ & 100--200 & 0.06$^{e}$  & 0.4$^{e}$  & 0.1$^{e}$ \\ 
             & W51~M                          & $\sim10^4$ & 100--200 & 0.007$^{e}$ & 0.07$^{e}$ & 0.1$^{e}$ \\ 
\hline
\end{tabular}
\end{center}
$^{a}$ Combining $N({\rm CH}_2)$ from \citet{lyu} with a hydrogen column density as quoted in \citet{vandishoeck86}.\\
$^{b}$ Based on $N({\rm CH})$=2.5$\times10^{13}$~cm$^{-2}$ from \citet{vandishoeck89}.\\
$^{c}$ Combining $N({\rm CH}_2)$ from \citet{lyu} with a hydrogen column density from \citet{1996ApJ...465..245S}.\\
$^{d}$ Based on $N({\rm CH})$=8$\times10^{13}$~cm$^{-2}$ from \citet{vandishoeck89}.\\
$^{e}$ Taking values as quoted by \citet{hollis} - but note that $X$(CH) could be significantly higher - see text.\\
\end{table*}

\subsection{\object{NGC7023}}

The ISO database also contains LWS FP observations of CH$_{2}$ towards three positions in the \object{NGC7023} PDR. However, no lines are detected above the noise in the spectra. This is consistent with LWS grating observations in which no OH, CH or CH$_{2}$ were detected toward any position in \object{NGC7023} \citep{fuente}.

\section{Discussion}

\subsection{Comparison with models}

In Table~\ref{comp} we summarise the results from several chemical models for diffuse \citep{vandishoeck86} and dense clouds \citep{lee}, as well as a recent model that takes both gas-phase and grain-surface chemistry into account \citep{ruffle}. In these models, the abundances of CH and CH$_{2}$ are generally predicted to be higher towards lower densities and temperatures. We also include the results from a PDR model of the \object{IC63} nebula from \citet{jansen95} - this is thought to have similar physical conditions to the \object{Sgr~B2} envelope ($n\sim10^5$~cm$^{-3}$; 650--900 times the average interstellar UV field). In this model, CH and CH$_{2}$ peak towards the edge of the PDR at A$_{\rm v}<2$, with [CH/CH$_{2}$] generally $\sim$2 (although in a model with higher fractional abundance of carbon they calculated a value $\sim$20). The CH abundance reached is relatively high compared to the dense cloud models with similar $n_{\mathrm{H}}$. In PDR models with higher density and more intense UV fields, even larger CH and CH$_{2}$ abundances can be reached in the outer layers \citep{sternberg}.
 
Table~\ref{comp} compares our estimated abundances with the models. In the line of sight clouds towards \object{Sgr~B2} ($-$100 to $+$30~km~s$^{-1}$), \citet{greaves96} have examined the abundances of 11 species observed at 3~mm (HCO$^{+}$, HCN, HNC, CN, CCH, C$_{3}$H$_{2}$, CS, SiO, N$_{2}$H$^{+}$, CH$_{3}$OH and SO) and found that overall, the chemistry in these features is similar to the dark cloud TMC1. However, our abundances are much closer to diffuse cloud values, showing that CH and CH$_{2}$ may reside in the less dense parts of the clouds. In \object{Sgr~B2} itself, our derived abundances are much more uncertain due to the difficulty in estimating $N({\mathrm{H}})$, but are lower than the line of sight clouds, consistent with denser material in \object{Sgr~B2}.

We can also compare our results with the previous observations of CH$_{2}$ in diffuse \citep{lyu} and dense clouds \citep{hollis}. Using the results of \citet{lyu} with hydrogen abundances for their lines of sight \citep[see][]{vandishoeck86,1996ApJ...465..245S}, we calculate value consistent with our line of sight values and the diffuse cloud model listed in Table~\ref{comp}. For the high densities and temperatures prevailing in the \object{Orion-KL} and \object{W51~M} hot cores, \citet{hollis} derive $X$(CH$_2) \approx 4\times10^{-9}$ and $7\times10^{-10}$, respectively, carefully considering beam-filling factors. These values are closer to our results for \object{Sgr~B2}.

\subsection{CH/CH$_2$ chemistry}

Both CH and CH$_{2}$ are formed by the dissociative recombination of CH$_{3}^{+}$ and primarily destroyed by reaction with atomic oxygen to form HCO, a constituent of many more complex molecules. \citet{hollis} compared their derived CH$_2$ abundance in the Orion-KL and W51~M hot cores with radio observations of CH to determine branching ratios for the formation of CH and CH$_{2}$ of 0.1 and 0.9 respectively. The actual branching ratios for the dissociative recombination of CH$_{3}^{+}$ have since been measured in the lab by \citet{vejby}. There are 4 important pathways, with CH$_{2}$ as the major product (40\%), two reactions producing CH (30\%) and the remainder producing atomic carbon (30\%). This appears to agree with models of diffuse and translucent clouds, which required a high abundance of CH$_{2}$ to reproduce the measured column densities of C$_{2}$ \citep{vandishoeck86,vandishoeck89}

The above discussion indicates that CH$_{2}$ should be more abundant than CH. However, our observations show that in the line of sight towards \object{Sgr~B2}, CH$_{2}$ has less than half the abundance of CH - if this were not the case, we would have detected much stronger CH$_{2}$ absorption. The observations towards \object{W49} and the non-detection towards \object{Sgr~A$^{\ast}$} also appear to confirm this. Furthermore, the results of \citet{lyu} towards HD154368 combined with previous measurements of CH (see Table~\ref{comp}) show very good agreement with the \object{Sgr~B2} ratio.

These results are contrary to the previous observations towards \object{Orion-KL} and \object{W51~M} where CH$_{2}$ was found to be 10 times more abundant than CH. This may be due to the difficulty in determining the CH column density equivalent to the measured CH$_{2}$ emission in these sources. \citet{hollis} used CH column densities derived by \citet{rydbeck}, giving $N(\rm{CH})=2\times10^{13}$~cm$^{-2}$ towards \object{W51~M}. In contrast, \citet{turner88} made a detailed study of the $\Lambda$-doublet satellite lines from the first rotationally excited state towards \object{W51}, giving a value for the 57~km~s$^{-1}$ velocity component that is more than 100 times larger: 6.2$\times10^{15}$~cm$^{-2}$. This indicates that [CH/CH$_{2}$] could be as high as 30, although the radio CH line observations are likely to sample a much larger and cooler region than the 70 GHz CH$_2$ lines which require high temperatures to be excited.

This shows it is extremely important to have consistent measurements of CH and CH$_{2}$ (with transitions at similar energies, using the same beam size and with consistent calibration) in order to be able to make a good comparison. So far the current data towards \object{Sgr~B2} provides the best comparison of the two species.

In order to explain the low abundance of CH$_{2}$ with respect to CH observed towards \object{Sgr~B2}, we require other formation/destruction processes to be important for the CH/CH$_{2}$ balance. The PDR model of \citet{jansen95} shows that the presence of strong UV radiation can reproduce our observed [CH/CH$_{2}$] ratio. This may explain the ratio in \object{Sgr~B2} but in the galactic spiral arm clouds, the UV field is much weaker and densities lower than used by \citet{jansen95}. \citet{viti_b} have modelled the hydrocarbon chemistry in diffuse and translucent clouds ($n$=300~cm$^{-3}$) including the interaction of gaseous C$^{+}$ with dust grains. This leads to the production of hydrocarbons by surface reactions. In some of their models (for $A_{\mathrm{v}}$=4), they find column densities approaching those that we find in the line of sight clouds towards \object{Sgr~B2} and [CH/CH$_{2}$] ratios 7.7--13.6. These surface reactions could explain our high observed [CH/CH$_{2}$] ratio.

\section{Summary}

We have made the first detection of the low-lying rotational transitions of the CH$_{2}$ molecule in the ISM towards \object{Sgr~B2} and \object{W49~N} in absorption. We do not detect CH$_{2}$ towards \object{Sgr~A$^{\ast}$} or \object{NGC7023}. For \object{Sgr~B2} we were able to compare these observations with measurements of the ground state rotational lines of CH, providing a good estimate of the total column density of both species along the line of sight. These observations provide the best comparison of the two species to date, giving a [CH/CH$_{2}$] ratio of 2.7$\pm$0.5, probably fairly constant in all the observed velocity components. The results towards \object{W49~N} appear to agree with this ratio, giving a lower limit of $\sim$1.0.

Comparison with chemical models shows that the abundances in the line of sight clouds are close to diffuse cloud conditions whereas in \object{Sgr~B2} itself they indicate denser gas is present. Our high [CH/CH$_{2}$] ratio can be explained by models including grain surface reactions \citep[e.g.][]{viti_b}.

Future observations in the FIR with SOFIA and in the sub-millimetre with telescopes such as the Atacama Pathfinder Experiment (APEX) will be highly interesting to extend the study of CH$_{2}$ to other sources.

\begin{acknowledgements}

We wish to thank T. W. Grundy (RAL) for supplying us with the LWS grating spectrum of \object{W49~N} processed using the latest 
version of the strong source correction and L01 post-processing pipeline. The LWS Interactive Analysis (LIA) package is a joint development of the ISO-LWS Instrument Team at the Rutherford Appleton Laboratory (RAL, UK - the PI Institute) and the Infrared Processing and Analysis Center (IPAC/Caltech, USA). The ISO Spectral Analysis Package (ISAP) is a joint development by the LWS and SWS Instrument Teams and Data Centres. Contributing institutes are CESR, IAS, IPAC, MPE, RAL and SRON.

\end{acknowledgements}

\bibliographystyle{aa}
\bibliography{1598bibl}

\appendix
\section{Observations used}

\begin{table}
\leavevmode
\footnotesize
\caption{\label{tdts}Log of the observations used.}
\begin{center}
\begin{tabular}[!t]{lcccc}
\hline
\hline
Transition & Observing  & ISO TDT      & LWS        & Resolution \\
           & Mode       & Number       & Detector   & (km~s$^{-1}$) \\
\hline 
\multicolumn{5}{l}{{\bf $^{{\bf 12}}$CH and $^{{\bf 13}}$CH  \object{Sgr~B2} } } \\
$J$=3/2--1/2  (149~$\mu$m)   & L03  & 50700208 & LW3 & 36    \\
 (\& para-CH$_{2}$ 3$_{12}$--3$_{03}$$^{a}$) & L03  & 84500102 & LW3 & 36    \\
                             & L03  & 50600603 & LW3 & 36    \\
                             & L03  & 50600814 & LW4 & 36    \\
                             & L03  & 50900521 & LW4 & 36    \\
$J$=3/2--5/2  (115~$\mu$m)$^{a}$  & L03  & 50601013 & LW2 & 34    \\
\hline 
\multicolumn{5}{l}{{\bf ortho-CH$_{{\bf 2}}$ \object{Sgr~B2}}}   \\
1$_{11}$--0$_{00}$ (127~$\mu$m) & L03  & 50700511 & LW2 & 34    \\
                                & L03  & 50800819 & LW3 & 34    \\
                                & L04  & 47600907 & LW2 & 34    \\
2$_{11}$--2$_{02}$ (153~$\mu$m)$^{a}$ & L03  & 50700208 & LW4 & 35    \\
                                      & L03  & 50800317 & LW4 & 35    \\
                                      & L03  & 50400823 & LW4 & 35    \\
                                      & L03  & 50600405 & LW3 & 35    \\
                                      & L03  & 83600317 & LW3 & 35    \\
3$_{13}$--2$_{02}$ (94~$\mu$m)  & L03  & 50800218 & LW1 & 34    \\
                                & L03  & 50400823 & LW1 & 34    \\
                                & L03  & 50700610 & LW1 & 34    \\
2$_{20}$--1$_{11}$ (50~$\mu$m)$^{a}$ & L03  & 50800218 & SW2 & 64    \\
\hline 
\multicolumn{5}{l}{{\bf para-CH$_{{\bf 2}}$ \object{Sgr~B2}}}   \\
1$_{10}$--1$_{01}$ (156~$\mu$m)$^{b}$  & L03  & 50700707 & LW4 & 35    \\
                                       & L03  & 50700610 & LW4 & 35    \\
                                       & L03  & 50800317 & LW4 & 35    \\
2$_{12}$--1$_{01}$ (107~$\mu$m) & L03  & 50800515 & LW2 & 32    \\
                                & L03  & 50800819 & LW2 & 32    \\
2$_{21}$--1$_{10}$ (51~$\mu$m)$^{a}$ & L03  & 50800216 & SW2 & 64    \\
                                     & L03  & 50600814 & SW2 & 64    \\
                                     & L03  & 50700511 & SW2 & 64    \\
                                     & L03  & 50900521 & SW2 & 64    \\
\hline 
\hline 
\multicolumn{5}{l}{{\bf CH$_{{\bf 2}}$ \object{W49~N}}} \\
1$_{11}$--0$_{00}$ (127~$\mu$m)       & L04  & 49900406 & LW2 & 34    \\
2$_{12}$--1$_{01}$ (107~$\mu$m)$^{a}$ & L04  & 49900406 & LW2 & 32    \\
\hline 
\multicolumn{5}{l}{{\bf CH \object{W49~N}}}   \\
$J$=3/2--1/2  (149~$\mu$m)            & L01  & 52700702 & LW4 & 1200   \\
\hline 
\hline 
\multicolumn{5}{l}{{\bf CH$_{{\bf 2}}$ \object{Sgr~A$^{{\bf \ast}}$}}}  \\
1$_{11}$--0$_{00}$ (127~$\mu$m)$^{a}$ & L03  & 50900204 & LW2 & 34    \\
2$_{12}$--1$_{01}$ (107~$\mu$m)$^{a}$ & L03  & 50800902 & LW2 & 32    \\
                                      & L03  & 85000139 & LW2 & 32    \\
                                      & L03  & 50900444 & LW1 & 32    \\
                                      & L03  & 85000247 & LW1 & 32    \\
\hline 
\multicolumn{5}{l}{{\bf CH \object{Sgr~A$^{{\bf \ast}}$}}}   \\
 $J$=3/2--1/2  (149~$\mu$m)           & L03  & 84900448 & LW3 & 34    \\
\hline
\hline
\multicolumn{5}{l}{{\bf CH$_{{\bf 2}}$ \object{NGC7023}} (3 different positions)}   \\
1$_{11}$--0$_{00}$ (127~$\mu$m)$^{a}$ & L04  & 33901603 & LW2 & 34    \\
                                      & L04  & 34601501 & LW2 & 34    \\
                                      & L04  & 56000208 & LW2 & 34    \\
\hline
\end{tabular}
\end{center}
$^{a}$ Not detected.\\
$^{b}$ Blended with NH line.
\end{table}

\end{document}